\documentclass[pre,twocolumn]{revtex4}

\usepackage{graphicx}
\usepackage{bm}
\usepackage{color}

\usepackage{amssymb, dsfont}

\newcommand{\R}{{\mathbf r}}

\newcommand{\f}{{\mathbf f}}

\newcommand{\xx}{{\mathbf x}}
\newcommand{\kk}{{\mathbf k}}
\newcommand{\yy}{{\mathbf y}}

\newcommand{\BEQ}{\begin{equation}}
\newcommand{\EEQ}{\end{equation}}
\newcommand{\BEA}{\begin{eqnarray}}
\newcommand{\EEA}{\end{eqnarray}}
\newcommand{\nn}{\nonumber}
\newcommand{\p}{\partial}

\newcommand{\beq}{\begin{equation}}
\newcommand{\eeq}{\end{equation}}
\newcommand{\bea}{\begin{eqnarray}}
\newcommand{\eea}{\end{eqnarray}}

\begin{document}

\title{Critical Phenomena in Active Matter}

\author{M.~Paoluzzi$^{1}$}
\email{mpaoluzz@syr.edu}
\author{C.~Maggi$^{2}$}
\author{U. Marini Bettolo Marconi$^{3}$}
\author{N.~Gnan$^{4}$}

\affiliation{
$^1$ Department of Physics, Syracuse University, Syracuse NY 13244, USA \\
$^2$Dipartimento di Fisica, Universit\`a di Roma ``Sapienza'', 
I-00185, Roma, Italy\\
$^3$Scuola di Scienze e Tecnologie, Universit\`a di Camerino, Via Madonna delle Carceri, 62032, Camerino, INFN Perugia, Italy\\
$^4$CNR-ISC, UOS Sapienza, P.le A. Moro 2, I-00185, Roma, Italy
}
\begin{abstract}
We investigate the effect of self-propulsion on a mean-field order-disorder transition.
Starting from a  $\varphi^4$  scalar field theory subject to an exponentially correlated noise, we exploit  the Unified  Colored Noise Approximation to map the non-equilibrium active dynamics onto an effective equilibrium one. This allows us  to follow the evolution of the second-order critical point as a function of the noise parameters: the correlation time $\tau$ and the noise strength $D$. Our results suggest that the universality class of the model remains unchanged.
We also estimate the effect of Gaussian fluctuations on the mean-field approximation finding an Ornstein-Zernike like expression for the static structure factor at long wave lengths. Finally, to assess the validity of our predictions, we compare the mean-field theoretical results with numerical simulations of active Lennard-Jones particles in two and three dimensions, finding a good qualitative agreement at small $\tau$ values.
\end{abstract}
\maketitle
\section{Introduction}
Motile cells, living bacteria, synthetic swimmers, flock of birds and school of fish are only a few examples of active systems able to give rise to a plethora of fascinating phenomena that spontaneously arise from their collective behavior \cite{Marchetti13,Cavagna14,Vicsek12}.  
In order to reproduce and understand the emergence of cooperative dynamics in active systems,  several-minimal models have been put forward, being mostly based on self-propelled agents, hydrodynamics theories but also on rule-base models with alignment interactions \cite{Cates12,Vicsek95,Chate08,Baskaran09}.
In spite of their minimal ingredients, these model systems display a highly collective behavior which results in large-scale pattern formations \cite{Redner13}, aggregation \cite{Tailleur08}, swarming \cite{Vicsek95}, off-equilibrium order-disorder transition \cite{Baskaran09}, peculiar rheological properties and disordered arrested states \cite{Berthier13,Henkes11,Bi15a,Bi15}.
Such a rich phenomenology shares many similarities with collective behavior in condensed matter physics where the emerging of a cooperative dynamics is intimately related  to the concept of phase transitions \cite{LeBellac91}.
The analogy between collective behavior in condensed matter and spontaneous aggregation in biological or synthetic systems, suggests that a coarse-grained procedure which neglects  the complexity of active agents could  reproduce, at least qualitatively, the observed phenomenology \cite{Cavagna14,Baskaran09}.  

Notable attempts in this direction have focused on specific models of isotropic self-propelled particles without aligning interactions \cite{Tailleur08,maggi2015,Farage15,Fodor16,Szamel15}. The fundamental ingredient that defines these non-equilibrium models is that the random force acting on each particle is not of thermal origin, i. e., is not a Brownian noise, but is a self-propulsion force that decorrelates on a time-scale $\tau$.
Early theoretical approaches 
were based on the idea of recasting the non-equilibrium dynamics in an effective equilibrium one with a density dependent diffusion coefficient \cite{Tailleur08,Cates12} suggesting a novel phase transition known as ``motility induced phase separation''.  Following the same idea, some of us have recently shown that the steady state distribution of  many active particles driven by Gaussian colored noise can be mapped onto an equilibrium problem where the noise amplitude and its correlation time play the role of control thermodynamic variables \cite{maggi2015,marconi2015,marconi16a,marconi16b}. 
In that study, the mapping to an effective equilibrium dynamics has been obtained thanks to the ``unified colored noise approximation'' (UCNA) \cite{Jung87,Hanngi95}. In addition, the  random driving forces have been modeled by an Ornstein-Uhlenbeck process (OUP) which gives rise to a self-propulsion that is Gaussian distributed and exponentially correlated in time. 
An exponentially correlated propulsion force characterizes also active Brownian \cite{Fily12,Farage15} and ``run and tumble'' dynamics \cite{Koumakis14}. The OUP has been shown to model quite well the behavior of passive tracers in active suspensions \cite{Wu00,Maggi14}.  Recently, many attentions are devoted by several research groups in modeling active particle systems by means of OUP \cite{Fodor16,Szamel16,Szamel15}.

Although the Gaussian colored noise model has been analyzed at the level of few particles~\cite{maggi2015}, in the case of a many particles system it presents the same insurmountable difficulties of the equilibrium many-body problem.

From this perspective, it would be desirable to develop a coarse-grained version of the model for studying phase transitions
especially to understand the role played by the memory of the noise on phase behavior. 
To this aim, in this article we propose and investigate a Gaussian colored-noise driven field theory based on the UCNA. 
In particular, we focus on the effect of colored noise on a second-order phase transition. 
In this framework,  we can compute the shift in the critical temperature due to the finite correlation time of the driving force.
The external parameter $\tau$ 
changes the location of the critical point but
not the universality class of the model. 
We find a reentrant behavior of the critical curve in the activity-noise phase diagram showing that, while for small value of $\tau$, phase transition is enhanced by the correlation time of the noise, at larger $\tau$ this tendency is inverted. Moreover, we compute the Gaussian fluctuations around the mean-field obtaining an Ornstein-Zernike (OZ) like expression for the static structure factor at low wave lengths. The OZ expression predicts a power-law divergence of the  correlation length at the critical point. The analytical mean-field predictions are compared with numerical simulations of a monodisperse active Lennard-Jones  fluid in two and three dimensions finding a good agreement at small $\tau$ values.

\section{The Model}
Critical phenomena are a special example of phase transitions and 
play a pivotal role in Statistical Mechanics \cite{Wilson93,ZJustin,LeBellac91}.
Landau Model is the common starting point to address a phase transition. 
In order to extend the Landau theory to
Active Systems, as a first step we have to fix the universality class of the problem.
Without loss in generality for the aim of this paper, we will look at a scalar field
theory.  
The scalar theory can be generalized to other universality class, i. e.,
we can include vectorial or tensorial fields with alignment interactions to study the
emerging of nematic order \cite{DeGennes74,Narayan07,Chate06,Baskaran08,Bertin13,Thampi14}.

We are interested in the case of a system close to the critical point and described 
by a scalar order parameter $\varphi(x)$,
e. g., the magnetization in the Ising ferromagnet, or the 
density difference $\rho_L- \rho_G$ 
in the gas-liquid phase transition.
The thermodynamics can be obtained by considering
the equilibrium solutions of the corresponding relaxation
dynamics \cite{Hohenberg77}. In the case of gas-liquid
transition, one should consider the Model B dynamics.
However, Model A and Model B share the same static properties
that are related to the Hamiltonian $H[\varphi(x))]$ as follows
\BEA\label{static}
\mathcal{F}(\beta)&=&-\frac{1}{\beta} \log Z \\ \nn 
Z&=&\int \mathcal{D}\varphi(x) \, e^{- \beta H[\varphi(x)]} \, ,
\EEA
where $\beta=T^{-1}$ and $T$ is the temperature \footnote{We use unit such that the Boltzmann constant $k_B=1$.}.
To obtain the Landau-Ginzburg (LG) theory we perform the saddle point
approximation in Eqs. (\ref{static}). The value $\varphi\!=\!\varphi_{SP}$
is given by the self-consistency equations
\BEQ\label{saddle2}
\left. \frac{\delta H}{\delta \varphi(x)}\right|_{SP} = 0 \, , \; \; \;  \left. \frac{\delta^2 H}{\delta \varphi(x)^2}\right|_{SP} > 0 \, \, ,
\EEQ
and the LG free energy is $H[\varphi_{SP}]$.

\subsection{Model A with Exponentially Correlated Noise}
In order to extend such a mean-field picture to the active counterpart
we start by considering the purely dissipative 
dynamics of a zero-dimensional $\varphi^4$ scalar field 
theory subjected to an exponentially correlated noise.
The equation of motion for the field $\varphi$ can be written
in term of an auxiliary variable $\theta$ that undergoes 
an OUP	
\BEA\label{field_dyn}
\partial_t \varphi(t) &=& -\frac{\partial H}{\partial \varphi} + \theta(t) \\ \nn
\partial_t \theta(t)   &=& -\frac{\theta(t)}{\tau} + \frac{D^{1/2}}{\tau}\eta(t) \, ,
\EEA
where the zero-mean noise $\eta$ is delta-correlated $\langle \eta(t) \eta(s)\rangle \!=\!2\delta(t-s)$,
and $D$ plays the role of (effective) temperature of the model.
The Hamiltonian $H$ is the standard $\varphi^4$
\BEQ\label{landau}
H[\varphi] =  \frac{a}{2} \varphi^2 + \frac{b}{4} \varphi^4 \, ,  
\EEQ
where $a$ depends linearly on $D$ and changes sign at $D_0$.
The $b$ coefficient is a positive constant. 
In the white-noise limit, that is recovered for $\tau\!\to\!0$, the 
steady-state solution of the Smoluchowski equation
associated to Eq. (\ref{field_dyn}) is the equilibrium
distribution function $P_{eq}[\varphi]\propto \exp{(- H[\varphi] / D})$. 

The stochastic differential equations (\ref{field_dyn}) can be  
rewritten as follows
\BEA\label{single}
\partial_t^2 \varphi &+& \frac{1}{\tau}\Gamma[\varphi] \,\partial_t \varphi=-\frac{1}{\tau}\frac{\partial H}{\partial \varphi} + \frac{D^{1/2}}{\tau}\eta(t) \, , \\ \nn
\Gamma[\varphi]&\equiv&1+\tau \frac{\partial^2 H}{ \partial \varphi^2}\,.
\EEA
The Unified Colored Noise Approximation is obtained neglecting
$\partial_t^2 \varphi$ in Eq. (\ref{single}) \cite{Jung87,Hanngi95}.
The corresponding Smoluchowski 
equation for $P[\varphi,t]$ reads \cite{ZJustin}
\BEQ
\partial_t P[\varphi,t] = \partial_{\varphi}  \left\{  \Gamma^{-1}[\varphi]  \left( D \partial_\varphi \Gamma^{-1}[\varphi]P[\varphi,t] -
\frac{\partial H}{\partial \varphi}  P[\varphi,t] \right) \right\}
\EEQ
and the steady-state distribution is
\BEA \label{ss}
P_{st}[\varphi]&=&\mathcal{N}e^{- H_{eff}[\varphi] / D} \\ \nonumber
H_{eff}[\varphi]&\equiv& H + \frac{\tau}{2}\left( \frac{\partial H}{\partial \varphi} \right)^2 - D \log | \Gamma[\phi] | \, .
\EEA
From Eq. (\ref{ss}) follows that $P_{st}$ has
the structure of an equilibrium distribution in terms of the effective Hamiltonian
$H_{eff}$.

\subsection{Critical line}
The critical line $D_c(\tau)$ is the curve along which the system undergoes
a second order phase transition. In a standard $\varphi^4$ theory, the location of
the critical point is determined by the coefficient of $\varphi^2$, i. e., the symmetry 
is spontaneous broken where $a$ changes sign.
In a mean-field model described by a LG free energy $\mathcal{F}_{LG}(\varphi)=a\varphi^2/2+B(\varphi)$,  the location of the critical point can be computed considering the equation \cite{ZJustin}
\BEQ
\left. \frac{\p^2}{\p \varphi^2} \mathcal{F}_{LG}\right|_{\varphi=0}= \left. a + \frac{\p^2}{\p \varphi^2} B\right|_{\varphi=0} =0 \, .
\EEQ
In our case, the LG free energy is $H_{eff}$, the expression for $D_c(\tau)$ is given by
\BEQ
\left. \frac{\p^2 H_{eff}}{\p \varphi^2} \right|_{\varphi=0}=0 \, .
\EEQ

Now we write
$a\!=\!a_0(D-D_0)$,  
with $a_0$ a positive constant.
The critical curve satisfies the equation
\BEQ\label{critical-1}
a(1+ a \tau)-\frac{6\tau b D}{1 +\tau a}=0\, \, ,
\EEQ 
and the only real and physical solution is
\BEA\label{critical}
D_c(\tau) &=& \frac{1}{\epsilon}\left[ \frac{\lambda}{\gamma}  + \gamma - \alpha \right] \\ \nn
\epsilon   &\equiv& 3 a_0^3\tau^2 \\ \nn
\lambda           &\equiv& a_0^4 \tau^2 + 18 a_0^3 b \tau^3 \\ \nn
\alpha      &\equiv&  2 a_0^2 \tau - 3 a_0^3 D_0 \tau^2 \\ \nn
\gamma   &\equiv&\left[ \frac{2 a_0^6 \tau^3 - 108 a_0^5 b \tau^4 + 162 a_0^6 b D_0 \tau^5 + \Delta}{2} \right]^{1/3} \\ \nn
\Delta        &\equiv& \left[ (2 a_0^6 \tau^3 - 108 a_0^5 b \tau^4 + 162 a_0^6 b D_0 \tau^5)^2  + \right. \\ \nn
&&\left. - 4 (a_0^4 \tau^2 + 18 a_0^3 b \tau^3)^3 \right]^{1/2} \, .
\EEA

\begin{figure}[!t]
\centering
\includegraphics[width=.5\textwidth]{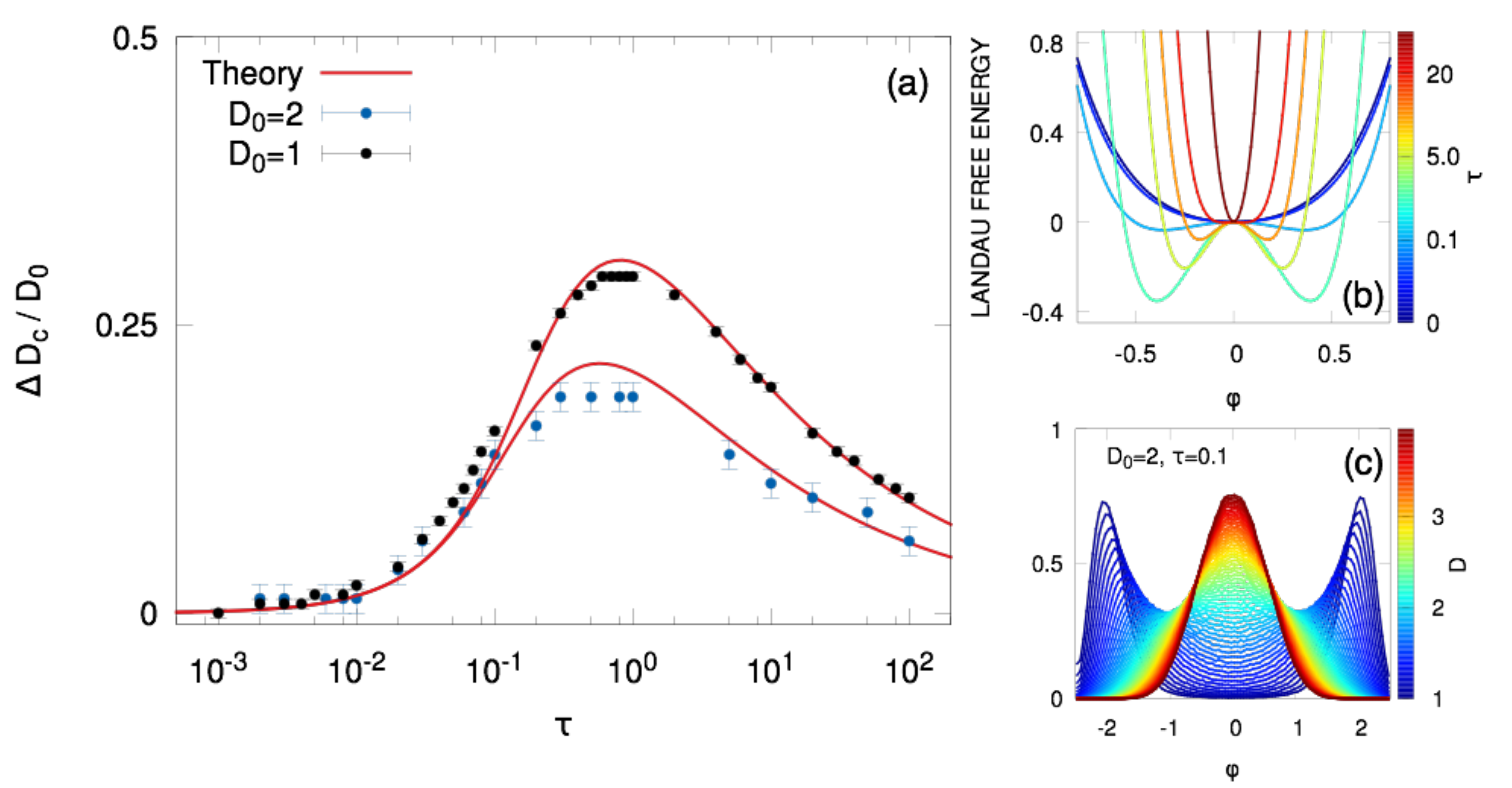}
\caption{Phase diagram in the activity-noise plane. 
Panel (a). The red curve is obtained through Eq. (\ref{critical}) with $\Delta D_c=D_c(\tau)-D_c(0)$, the symbols are obtained
via numerical integrations of Eqs. (\ref{field_dyn}).
Panel (b). The symmetric phase at high $D$ is represented by the quadratic free energy,
the spontaneous symmetry breaking phase at low $D$ by the double-well.
Panel (c). $P[\phi]$ obtained through numerical integration of Eqs. (\ref{field_dyn}).
}
\label{fig:fig1}      
\end{figure}

$D_c(\tau)$ increases for small $\tau$, reaches its maximum
value $D^*$ at $\tau^*$, and decreases to $D_0$ for $\tau\to\infty$.
As a consequence the phase diagram in the $(\tau,D)$ plane is reentrant.
This is shown in the left inset of Fig. (\ref{fig:fig1}) where LG free energy
is plotted increasing $\tau$ for $D_0\!<\!D\!<\!D^*$. The LG free energy develops 
a double well for $\tau_{-}\!<\!\tau\!<\!\tau_{+}$ (magenta and yellow curves),
with $\tau_{\pm}$ the solutions of $D\!=\!D_c(\tau)$. For $\tau<\tau_-$
or $\tau\!>\!\tau_+$ the system is in the symmetric phase (blue and red curves, respectively).

It is worth noting that a reentrant behavior of the Boyle's line 
has been observed in the virial series
of many mutually interacting particles in the presence of correlated noise \cite{marconi2015}. 
The small $\tau$ behavior indicates that memory effects in the dynamics
raise the critical effective temperature, suggesting that  the activity
promotes criticality.

In order to compare the analytical expression for $D_c(\tau)$ with the true order parameter dynamics,
we have solved numerically the non-equilibrium dynamics. 
Eqs. (\ref{field_dyn}) have been numerically integrated for $N_t\!=\!10^{6}$ steps with $\Delta t \!=\! 10^{-3}$. The parameters of the model are $a_0\!=\!4$, $D_0\!=\!1,2$ and $b\!=\!1$. 
From the trajectories $\varphi(t)$ we have computed $P[\varphi]\!=\!\langle \delta\left[ \varphi(t) - \varphi\right]\rangle_{t,\varphi(0)}$, where the angular bracket indicates both the averages, over the trajectories and over the initial condition. The critical point has been obtained by fitting the histogram of $P[\varphi]$ to $f(x)\!=\!A\exp{( -\tilde{a}\varphi^2-\tilde{b}\varphi^4)}$.
We have considered average over $5\cdot 10^2$ initial conditions. 
The resulting $P_{st}[\varphi]$ for $\tau\!=\!0.1$ is shown in Fig. (\ref{fig:fig1}), right inset.
As one can see in the main panel of the same figure, the theoretical curve $\Delta D_c/D_0$,
with $\Delta D_c=D_c(\tau) - D_c(0)$,
reproduces very well the numerical data in a wide range of $\tau$. 

For small $\tau$ we can approximate $\log{\Gamma(\varphi)}\sim D\tau\partial^2_\varphi H$ 
obtaining an effective $\varphi^6$ theory. It is well known in literature that $\varphi^6$ theory admits a tricritical point where the second order phase transition changes in a first order phase transition \cite{ZJustin}. However, in our model the tricritical point is located in an unphysical region. The effective hamiltonian reads
\BEA\label{phi6}
H_{eff}^{small}[\varphi] &=& \frac{\tilde{a}}{2} \varphi^2 + \frac{\tilde{b}}{4}\varphi^4 + \frac{\tilde{c}}{6}\varphi^6 \\ \nn
\tilde{a} &\equiv& a (1 + a \tau) -6 D b \tau \\ \nn
\tilde{b} &\equiv& b + 4 a b \tau \\ \nn
\tilde{c} &\equiv& 3 b^2 \tau \, .
\EEA
In this case, the critical line $D_c^{small}$ is given by $\tilde{a}=0$ and satisfies 
\BEQ\label{c6}
a(1+\tau a) - 6 D \tau b = 0 \, .
\EEQ
Along $D_c^{small}$, $\varphi_0$ behaves like $\varphi_0\!\sim\!(D-D_c)^{\beta}$
with $\beta\!=\!1/2$, i. e., the classical mean-field value for the $\beta$ exponent \cite{LeBellac91}.

\begin{figure*}[!th]
\centering
\includegraphics[width=1.\textwidth]{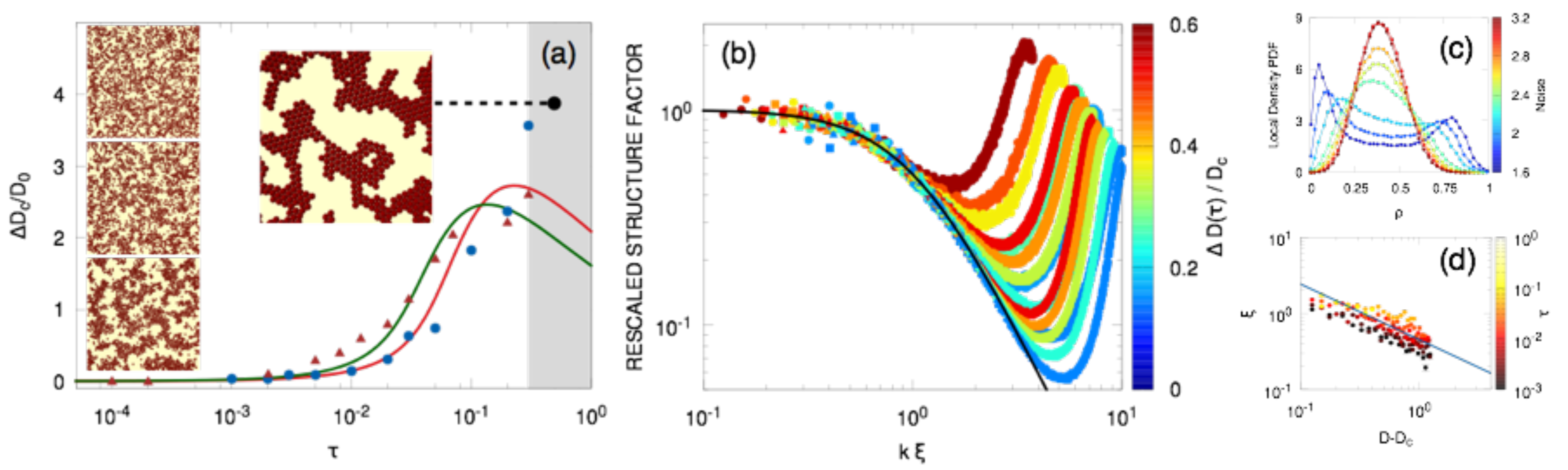}
\caption{Numerical simulations.
The symbols in panel (a) are the critical point of the active Lennard-Jones
fluid in two and three dimensions, blue circles and brown triangles, respectively. 
The full lines are the fit of the data with Eq. (\ref{critical}). 
Left snapshots: two-dimensional numerical simulations approaching $D_c(\tau)$ (from top to down) for $\tau=3\cdot 10^{-3}$.
Cluster crystallization occurs in the gray region.
Right snapshot: crystal clusters for $\tau=1.0$. 
In panel (b) it is shown the rescaled structure factor $S(k\xi)$
in two dimensions for $D>D_c$ and $\tau=10^{-3},10^{-2},10^{-1}$(circles, squares, and triangles). Symbols represent numerical data, the black curve is obtained
by fitting to a Lorentzian distribution. 
Panel (c). $P_\ell(\rho)$ in three dimensions for $\tau=3\cdot10^{-2}$ and $N_b=13$, the lines are a guide to the eye.
Panel (d). $\xi$ obtained from fitting $S(k)$ for small $k$ to $A(1+(\xi k)^2)^{-1}$.
The blue line is determined from fitting $\xi$ with $(D-D_c)^{-\nu}$.
}
\label{fig:fig2}       
\end{figure*}
%

\section{Numerical Simulations} \label{num}
Now we compare the mean-field picture with its
finite-dimensional counterpart. With this aim, we have performed numerical simulations of 
$N$ spherical particles interacting through 
$\phi(r)=4 e_0 ((r/\sigma)^{-12} - (r/\sigma)^{-6})$ in two ($d=2$) and three ($d=3$) dimensional
boxes of side $L$ with periodic boundary conditions. The density
of the system is $\rho_m=N/L^d$. The energy is measured in unit $e_0$ and the
density in unit $\sigma^{-d}$.
The microscopical model undergoes a gas-liquid (GL) phase separation
that is described by a conserved scalar order parameter $\rho_G - \rho_L$. The nature of the order parameter 
implies Model B dynamics along the phase separation \cite{Hohenberg77}.
However, the location of the critical point, i. e., the endpoint of the phase separation,  is a static property of the system.
As mentioned earlier, Model A and Model B show the same static properties:  
we will adopt the mean field scenario emerging from Model A to capture 
qualitatively the behaviour of the critical line in finite dimension. 

The self-propulsion is modeled by means of a random
driving force exponentially correlated in time.
The equation of motion of the particle $i$, with
$i\!=\!1,...,N$, is 
\BEQ
\dot{\R}_i\!=\! \f_i - \mu \sum_{j\neq i}\phi^\prime(r_{ij})\R_{ij}/r_{ij}  \, ,
\EEQ
where $\mu\!=\!1$ is the mobility and $r_{ij}\!\equiv\!|\R_i-\R_j|$. The random force satisfies $\langle f^\alpha_i \rangle\!=\!0$
and $\langle f^\alpha_i(t) f^\beta_j(s) \rangle\!=\!2D\delta_{ij} \delta_{\alpha\beta}e^{-|t-s|/\tau}/\tau$, where
the Greek symbols indicate the cartesian components.
In this picture the external parameters 
$D$ and $\tau$ can be independently varied as well as in the Landau model previously considered. 
For $\tau\!=\!0$ one recovers the Lennard-Jones fluid in the Brownian regime.
We study the system close to the LJ critical density $\rho_c\sim0.4$ ($N\!=\!2500$ in $2$d and $N\!=\!8000$ in $3$d). 
Moreover, performing simulations at different values of $\rho_m$ \footnote{ In two dimensions $\tau\!=\!0,10^{-3},3\cdot 10^{-3},5\cdot 10^{-5},10^{-1},3 \cdot 10^{-1},5 \cdot 10^{-1},6 \cdot 10^{-1},0,1,0.3,0.5$. In three dimensions $\tau\!=\!0,10^{-4},2\cdot 10^{-3},5 \cdot 10^{-3}, 8 \cdot 10^{-3},10^{-2}, 2 \cdot 10^{-2}, 3 \cdot 10^{-2}, 5 \cdot 10^{-2}, 7 \cdot 10^{-2}, 0.3,1,10$. We have explored $\rho_m\in[0.2,0.7]$}, we have checked that the value $\rho_c$ does not vary with $\tau$. 
The critical values $D_c(\tau)$ have been evaluated looking at the intersection points 
of the Binder cumulant $U_\ell=1-\langle \delta \rho^4 \rangle_\ell  / 3 \langle \delta \rho^2 \rangle_\ell^2$ \cite{Binder81} at different $\ell$,
with $\delta \rho = \rho - \langle \rho \rangle$,
where the average is defined as $\langle \mathcal{O} \rangle_\ell=\int d\rho P_\ell(\rho) \mathcal O (\rho)$. 
The block density distribution function $P_\ell(\rho)\equiv \langle \delta(\rho - \rho_k) \rangle$ is obtained by dividing the simulation box with linear size $L$ in $N_b$ cells of size $\ell=L/N_b$ and coarse-grained density $\rho_k$ with $k=1,...,N_b$\cite{Rovere88,Rovere89}. The behavior of $P_\ell(\rho)$ in three dimensions approaching the
transition is shown in Fig. (\ref{fig:fig2})-a (top inset).
In order to evaluate the intersection of $U_\ell$ as a function of $D$, we have performed simulations of $52$ different $D$ for each value of $\tau$. 

The resulting
phase diagram is shown in Fig. (\ref{fig:fig2})-a with
snapshots of the $2d$ simulations (bottom left inset in the same panel).
The full lines
are obtained by fitting the data with Eq. (\ref{critical}) leaving both $a_0$ and $b$
as free parameters. 
%
As one can see for small $\tau$ the theory reproduces quite well the numerical data. 
However with the model simulated we cannot probe the regime at larger  $\tau$ values since crystallization occurs at $\tau > 0.3$ (the gray area in Fig. (\ref{fig:fig2})-a). In order to prevent crystallization, one can introduce frustration
in the microscopical model considering, for instance, a binary mixture \cite{Kob94}.
Hence, the existence of the reentrance  in the activity-noise phase diagram remains an open question that we aim at answering in a future work.
It is worth noting that our starting point is a $\varphi^4$ theory. In such a case we can not
describe a phase diagram that shows both gas/liquid and gas/crystal phase
transition. Nevertheless, it is possible to generalize our mean-field model 
considering a different field theory in order to take into account other kind of phase transitions \cite{LeBellac91}. 

In the crystal regime, the nucleated liquid droplets rearrange into small crystal
clusters. The presence of the crystal clusters is evident in the snapshot shown in
Fig. (\ref{fig:fig2})-a, right.

\section{Gaussian Fluctuations}
Let us discuss the effect of the correlated noise on the Gaussian fluctuations around the mean field \cite{LeBellac91}.
The Hamiltonian in $d-$dimensions is
\BEQ\label{gauss}
H_{G}[\varphi(\xx,t)]=\frac{1}{2}\int d^dx\, \left[ (\nabla \varphi(\xx,t))^2 + a \varphi(\xx,t)^2 \right] \,.
\EEQ 	
We will consider both Model A/B dynamics with exponentially correlated noise. 
We can rewrite Eqs. (\ref{field_dyn}) including the spatial dependency in a compact way as follows
\BEA \nn
\partial_t \varphi(\xx,t) &=& - \left( i \nabla \right)^{2 \psi} \left( \frac{\delta H_{G}}{\delta \varphi(\xx,t)} \right) + \left( -\nabla \right)^\psi \theta(\xx,t) \\ 
\partial_t \theta(\xx,t)   &=& -\frac{\theta(\xx,t)}{\tau} + \frac{D^{1/2}}{\tau} \eta(\xx,t) \, .
\EEA
The exponent $\psi$
is $0$ (Model A) or $1$ (Model B). The noise is white $\langle \eta(\xx,t) \rangle=0$, and 
delta-correlated, $\langle \eta(\xx,t) \eta(\yy,s) \rangle = 2 \delta(\xx-\yy) \delta(t-s)$.

Now we introduce the spatial Fourier transform of a field $\phi(\xx)$ as
\BEQ
\phi_\kk=\frac{1}{(2 \pi)^d}\int d^dx \, e^{-i \xx \cdot \kk} \phi(\xx)\, .
\EEQ
We can perform the spatial Fourier transform of $\varphi(\xx,t)$, $\theta(\xx,t)$, and $\eta(\xx,t)$ obtaining
the time evolution of the $k-$th Fourier component of $\varphi$ in UCNA, i. e., considering $\partial_t^2 \varphi_k=0$, that is governed by the following equation 
\BEA\label{motion}
\partial_t \varphi_k &=& -\Gamma_k \varphi_k + \mathcal{D}_k \eta_k \\ \nn
\Gamma_k &\equiv&  k^{2 \psi} \frac{k^2 + a}{1 + \tau (k^2 + a)} \\ \nn
\mathcal{D}_k &\equiv& (i k)^{\psi} \frac{D^{1/2}}{1 + \tau (k^2 + a)} \, ,
\EEA
where $k\equiv | \kk |$.
Again, the noise $\eta_k$ is white $\langle \eta_k \rangle \!=\! 0$, and delta-correlated 
$\langle \eta_k(t) \eta_q(s)\rangle\!=\!2 \delta_{kq}\delta(t-s)$. 
We can estimate the critical slowing down exponent by averaging Eq. (\ref{motion})
over the noise. It follows that $\langle \varphi_k (t)\rangle \propto\exp{(-t/\tau_k)}$,
with $\tau_k\!=\!\Gamma_k^{-1}$.
Introducing $\xi^2\!=\!a^{-1}$, we can write the relaxation time of the mode $k$ as $\tau_k\!=\!\xi^z f(k \xi,\tau \xi^{-2})$.
The value of the dynamical critical exponent $z\!=\!2(1+\psi)$ turns out to be the same as in the case of equilibrium dynamics \cite{Hohenberg77}.

From Eq. (\ref{motion}) we can compute the stationary fluctuations
$\langle | \varphi_k |^2 \rangle = \xi^2 g(k\xi,\tau)$ with 
\BEQ\label{fluct}
g(k\xi,\tau)=\frac{D}{\left(k^2 \xi^2 + 1 \right) \left[ 1 + \tau (k \xi)^{2\psi} \xi^{-2(1+\psi)} (k^2 \xi^2 + 1) \right]} \, .
\EEQ
When $k\to0$, $\langle | \varphi_k |^2 \rangle$ diverges as $\xi=a^{-\nu}$, with the classical value $\nu=1/2$.
From Eq. (\ref{fluct}) follows that, above the transition, the static structure factor $S(k)$ for small $k$ is
well described by the usual OZ expression $S(k)\sim (k^2\xi^2 + 1)^{-1}$.  
In Fig. (\ref{fig:fig2})-b we show the rescaled $S(k)$ in $2d$ 
for $\tau=10^{-3},10^{-2},10^{-1}$ and $D>D_c(\tau)$. 
According to OZ, the rescaled $S(k)$ overlaps on the same master curve. 
Moreover, as it is shown in the inset of Fig. (2)-b, $\xi$ follows a power law $(D-D_c)^{-\nu_N}$ 
with $\nu_N=0.73$ for $N=2500$. 
The value of the exponent does not depend on $\tau$, i. e., as predicted by the mean-field picture.
Since we are in finite dimension, it is different from the classical value $1/2$ \cite{LeBellac91}.
It is worth noting that $\nu$ has been evaluated for $N=2500$. In order to estimate
the critical exponent $\nu$ in the thermodynamic limit, we should take into account the finite-size correction to $D_c$, i. e., $D_c(L)=D_c^\infty+a L^{-b}$ \cite{Palassini99}. This aspect will be investigated in a future work.

\section{Conclusions}
In this article, we have posed a fundamental question about the influence of self-propulsion
on the order-disorder transitions. 
By means of UCNA, we have recast the non-equilibrium dynamics 
of the model A/B in the presence of correlated noise onto an effective equilibrium theory.
In this way we have extended the static picture of the Landau theory of critical phenomena 
to order-disorder transitions in the presence of exponentially correlated noise.
The location of the critical point is a non-universal quantity because depends on the 
correlation time of the noise $\tau$. Through the effective equilibrium theory, we have
computed analytically the critical line $D_c(\tau)$, i. e., the shift in critical temperature due to the activity. 
Considering the numerical solution of the non-equilibrium dynamics of the order parameter in zero dimension, we have 
observed a good agreement between $D_c(\tau)$ and the critical points obtained numerically.
Moreover, the effective theory suggests that the out-of-equilibrium dynamics does not change the universality
class. This finding is in agreement with previous studies on Ising-like nonequilibrium models \cite{Mendes98,Oliveira92,Miranda87} where it has been observed that the 
absence of detailed balance on the microscopic scale does not change the universality class of the Ising model.

By performing numerical simulations in $2d$ and $3d$
of active LJ fluid driven by OUP, we have obtained that the mean-field scenario can be used to describe
the behavior of the critical line in the small $\tau$ regime. 
However, the reentrant behaviour predicted by the mean-field scenario does not occur in
the considered microscopical model. In particular, the numerical simulations show a gas/crystal 
phase transition at larger $\tau$ that can not be captured by the theory.

With the numerical data presented in this paper we can conclude that,
for small $\tau$ and independently on 
density, the {\itshape out-of-equilibrium} dynamics gives rise to a second order phase transition that shares
the same properties with its equilibrium counterpart, i. e., $\tau=0$.

Finally, we have evaluated the Gaussian fluctuations approaching the critical point from the disordered phase.
According to the theory,
we have demonstrated that the static structure factor at low $k$ is well described by OZ expression.  
Here, we have considered a $\varphi^4$ scalar field theory, however
our approach can be generalized to other field theories in order to study the properties of 
different universality classes under the effect of self propulsion.

\section*{Acknowledgments}
We thank M Cristina Marchetti for her critical reading of the early version of the manuscript.
MP was supported by the Simons Foundation  Targeted Grant in the Mathematical Modeling of Living Systems Number: 342354 and by the Syracuse Soft Matter Program.
CM was supported by the European Research Council under the European Union's
Seventh Framework Programme (FP7/2007-2013)/ERC grant
agreement no. 307940. NG acknowledges support from MIUR
(``Futuro  in  Ricerca"  ANISOFT/RBFR125H0M).


\end{document}